\documentstyle[11pt,newpasp,twoside,epsf]{article}

\begin{document}
\title{Global Evolution Models and the Extragalactic Background Light}
\author{S. Michael Fall}
\affil{Space Telescope Science Institute, 3700 San Martin Drive, Baltimore,
MD 21218, USA}

\begin{abstract}
This article outlines a method to interpret the extragalactic 
background light in terms of the large-scale, average properties 
of galaxies, including the comoving densities of stars and
interstellar gas, metals, and dust. 
These quantities are related by a series of coupled conservation-type
equations analogous to the equations of galactic chemical evolution.
This approach enables us to combine observations of the emission and 
absorption in galaxies and thus to relate their average stellar and 
interstellar contents.
Applications of the method include predictions of the global history 
of star formation from absorption-line observations and corrections 
to the cosmic UV emissivity for absorption by dust.
\end{abstract}

\section{Introduction}

The extragalactic background light (EBL) is an inherently average or
``global'' property of the present-day Universe.
It is a direction-averaged, time-integrated record of the emission and 
absorption of photons since recombination ($z \la 1000$).
As such, the EBL is closely related to the global history of star 
formation and, potentially, that of other processes, such as the 
accretion of matter onto black holes. 
The global history of star formation in turn is closely related to the
global histories of gas consumption and metal production in galaxies.
The purpose of this article is to explore some of these relations.
In this approach, one largely ignores the individuality of galaxies
and their internal complexity.
The primary focus, instead, is on the average properties of galaxies.

The global approach may be contrasted with others in which the 
individuality and diversity of galaxies are regarded as paramount.
In the latter, one attempts to predict, or at least to describe, the 
full variety of evolutionary paths followed by galaxies of different 
types, masses, and other properties.
This is necessary to interpret luminosity functions and number counts.
The EBL and other global quantities are then obtained by integrating 
over the these functions.
This approach is more challenging and perhaps more fundamental than 
the global approach in that it ties the interpretation of the EBL to 
a fairly complete understanding of the formation and evolution of 
galaxies.
The global approach forfeits some of this insight but is much simpler.
That is both its main limitation and its main advantage.

\section{Global Evolution: Overview}

One of the grand themes to emerge in the last few years is the idea 
that we may be able to determine the global histories of star formation, 
gas consumption, and metal production in galaxies from high redshifts 
to the present. 
This idea is illustrated in Figure~1, where we sketch the evolution of 
the contents of a large, comoving box. 
We can conveniently quantify the masses of the different constituents of 
the box by the corresponding mean comoving densities normalized to the 
present critical density. 
We are especially interested in the comoving densities of stars, gas 
(both inside and outside galaxies, i.e., ISM and IGM), metals, dust, 
and black holes: $\Omega_s$, $\Omega_{ism}$, $\Omega_{igm}$, $\Omega_m$, 
$\Omega_d$, and $\Omega_{bh}$, respectively. 
We are also interested in the cosmic emissivity $E_\nu$, the power 
radiated per unit comoving volume per unit rest-frame frequency $\nu$, 
and the EBL intensity $J_\nu$, the power received per unit solid angle 
of sky per unit area of detector per unit observed frequency $\nu$.

\begin{figure}
\plotfiddle{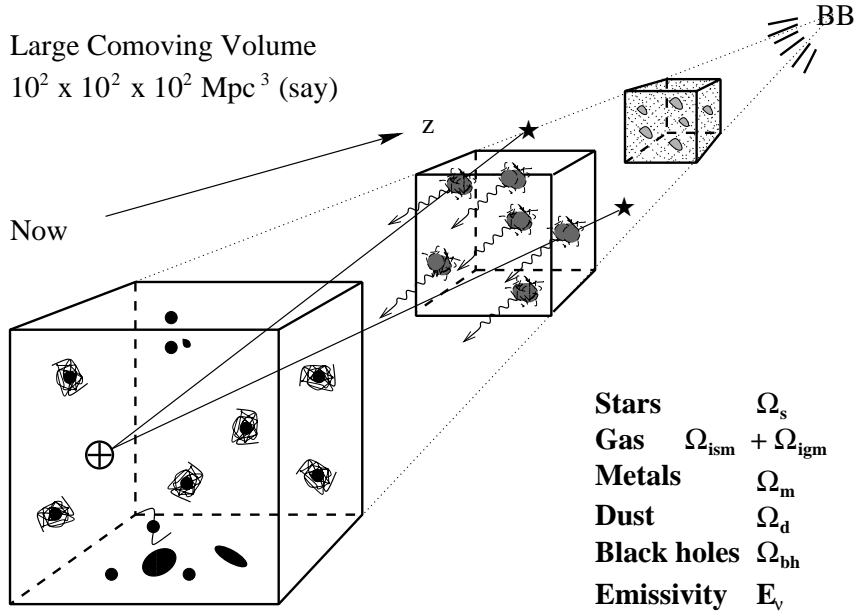}{3.5in}{0}{50}{50}{-165}{5}
\caption{Evolution of the contents of a large comoving volume of the 
Universe, from the Big Bang to the present, including galaxies and the 
IGM.
The wavy lines represent the light emitted by stars, AGN, and dust in 
galaxies; the straight lines represent the light emitted by quasars and
then partially absorbed or scattered in the ISM of foreground galaxies
and the IGM.}
\end{figure}

After recombination, our comoving box is filled with neutral, metal-free 
gas with nearly uniform density. 
Perturbations in this intergalactic medium (IGM) eventually condense, 
probably by gravitational clumping and inflow, into protogalaxies. 
Stars then form in the resulting interstellar medium (ISM). 
They produce metals and may drive outflows of gas from galaxies. 
In this way, both the ISM and IGM may be enriched with metals. 
Some of the metals remain in the gas phase; others condense into solid 
dust grains. 
Black holes form in the nuclei of some, perhaps even all, galaxies and, 
when fueled, can power active galactic nuclei (AGN). 
Some of the radiation emitted by stars and AGN propagates freely, while 
the rest is absorbed and then emitted at longer wavelengths by dust. 
Thus, the radiation we detect from galaxies tells us primarily about their 
star, AGN, and dust contents.
The spectra of high-redshift quasars contain signatures of the absorption 
and scattering of radiation by the intervening ISM and IGM (absorption 
lines, reddening, etc). 
Such observations tell us primarily about the composition and comoving 
densities of the ISM and IGM.

Exactly how all this happens is not yet known, of course. 
It should be clear from Figure~1 and the commentary above, however, 
that the constituents of our comoving box, including the radiation 
that propagates through it, are very much interrelated. 
In fact, the corresponding comoving densities must obey a series of 
coupled conservation-type equations. 
Clearly, $\Omega_s$, $\Omega_{ism}$, and $\Omega_{igm}$ must add up 
to $\Omega_{\rm baryon}$, a constant. 
Similarly, $\Omega_m^s$, $\Omega_m^{ism}$, and $\Omega_m^{igm}$, the 
comoving densities of metals in stars, the ISM, and the IGM, must add 
up to $\Omega_m$. 
Moreover, $\Omega_m$ remains a fixed fraction of $\Omega_s$ on the 
assumption that the global yield is constant and that delayed recycling 
is negligible (a good approximation in the present context). 
The cosmic emissivity $E_\nu$ depends on the star formation rate
$\dot\Omega_s$, black hole fueling rate $\dot\Omega_{bh}$, and the 
amount of reprocessing by dust and hence $\Omega_d$. 
The EBL intensity is related to the cosmic emissivity by a radiative 
transfer equation; for wavelengths at which the IGM is transparent, 
this gives
\begin{equation}
J_{\nu} = {c\over4\pi} \int^{t_0}_0 E_{(1+z)\nu} dt,
\end{equation}
where $t_0$ is the present age of the Universe.
It is likely that AGN make relatively small contributions to 
$E_\nu$ and $J_\nu$ and can be neglected in a first approximation.

\section{Global Evolution: Equations}

To obtain a set of equations for the evolution of the average stellar
and interstellar contents of galaxies, we begin with the familiar
equations of galactic chemical evolution.
In this approach, the IGM is regarded as a reservoir with which a
galaxy can exchange material by inflow or outflow.
Following standard practice, we denote the mass of stars and ISM 
(gas and dust) in a galaxy by $M_s$ and $M_g$ ($\equiv M_{ism}$), 
the interstellar metallicity by $Z$ ($\equiv M_m^{ism}/M_{ism}$), 
the inflow or outflow rate by $\dot M_f$, the metallicity of 
inflowing or outflowing material by $Z_f$, and the nucleosynthetic 
yield of heavy elements, averaged over the stellar initial mass 
function, by $y$. 
In the approximation of instantaneous recycling and $Z \ll 1$, these 
quantities are related by 
\begin{equation}
{d\over{dt}}(M_g + M_s) = \dot M_f, 
\end{equation}
\begin{equation}
{d\over{dt}}(Z M_g) = y{d\over{dt}}M_s - Z{d\over{dt}}M_s + Z_f\dot M_f.
\end{equation}
Equation~(3) gives the total rate of change of the mass of metals in
the ISM; on the right-hand side, the first term is the gain by stellar 
production and ejection (proportional to the rate of star formation since 
most of the metals are synthesized in short-lived stars), the second term
is the loss by removal of the ISM (including its metals) by forming stars, 
while the third term is the gain or loss by exchange with the IGM (see 
Tinsley 1980 or Pagel 1997 for complete explanations and derivations).

If we now sum equations~(2) and~(3) over all galaxies in a large comoving
volume and then divide by that volume and the present critical density,
we obtain what are sometimes called the equations of cosmic chemical 
evolution:
\begin{equation}
{d\over{dt}}(\Omega_g+\Omega_s) = \dot\Omega_f, 
\end{equation}
\begin{equation}
{d\over{dt}}(Z\Omega_g) = y_e{d\over{dt}}\Omega_s - Z{d\over{dt}}\Omega_s 
+ Z_f\dot\Omega_f.
\end{equation}
Here, the effective yield $y_e$ depends on the nucleosynthetic yield
$y$ and the spread in the interstellar metallicities of galaxies weighted
by their star formation rates. 
In the hypothetical case that all galaxies have the same interstellar 
metallicity, $y_e$ is the same as $y$.
In general, the two yields will differ, but this is not a serious problem
because one can adjust $y_e$ so that the mean metallicity in the models 
agrees, at some particular time, such as the present, with the observed
value.
This is the same approach taken in nearly all models of galactic 
chemical evolution.
Indeed, if one were modeling the evolution of a chemically inhomogeneous 
galaxy, the same issues would arise in the derivation and solutions of 
equations~(2) and~(3); the parameter $y$ would then have to be regarded 
as an effective yield, not equal to the nucleosynthetic yield.

\section{Absorption-Line Systems}

The average interstellar properties of galaxies can be determined from 
the statistics of quasar absorption-line systems as follows.
Let $f(N_x,z)$ be the column density distribution of particles of any
type $x$ that absorb or scatter light.
These might, for example, be hydrogen atoms ($x=$~HI), metal ions ($x=m$), 
or dust grains ($x=d$). 
By definition, $H_0 (1+z)^3 |dt/dz| f(N_x,z)dN_xdz$ is the mean number 
of absorption-line systems with column densities of $x$ between $N_x$ and 
$N_x+dN_x$ and redshifts between $z$ and $z+dz$ along the lines of sight 
to randomly selected background quasars.
These lines of sight are very narrow (the projected size of the continuum
emitting regions of quasars, less than a light year across) and pierce 
the absorption-line systems at random angles and impact parameters.
One can show that the mean comoving density of $x$ is given by 
\begin{equation}
\Omega_x(z) = {8 \pi G m_x \over 3 c H_0} \int_0^\infty N_x f(N_x,z) dN_x,
\end{equation}
where $m_x$ is the mass of a single particle (atom, ion, or grain).
Equation~(6) plays a central role in this subject.
It enables us to estimate the mean comoving densities of many quantities
of interest without knowing anything about the structure of the 
absorption-line systems.
In particular, we do not need to know their sizes or shapes, whether they
are smooth or clumpy, and so forth.
A corollary of equation~(6) is that the global interstellar metallicity, 
$Z\equiv\Omega_m^{ism}/\Omega_{ism}$, is given simply by an average over 
the metallicities of individual absorption-line systems weighted by their 
total column densities.

The absorption-line systems of most interest in the present context 
are the damped Ly$\alpha$ (DLA) systems. 
It is widely believed that they constitute the ISM of galaxies and 
protogalaxies and hence are the principal sites of star formation in 
the Universe (Wolfe et al. 1986).
There are excellent reasons to adopt this as a working hypothesis. 
First, the DLA systems have, by definition, $N_{\rm HI}\ga
10^{20}$~cm$^{-2}$, and this, at least at low redshifts, is near or
slightly below the threshold for star formation (Kennicutt 1989).
Second, the DLA systems contain at least 80\% of the HI in the Universe 
and appear to be mostly neutral.
The other absorption-line systems, those with $N_{\rm HI}\la
10^{20}$~cm$^{-2}$, probably contain more gas in total than the DLA
systems, but this must be diffuse and mostly ionized.
In the following, we regard non-DLA systems as belonging to the IGM,
even though some of them might actually be associated with the outer,
tenuous parts of galaxies.
This distinction---between the mostly-neutral ISM, where stars form, and
the mostly-ionized IGM, where they do not---is certainly valid at the 
present epoch (Zwaan et al. 1997).
Thus, the DLA systems are often referred to as DLA galaxies.
The precise nature of these objects---whether they are large or small, 
disk or spheroid---remains to be determined, probably by direct imaging
(Le~Brun et al. 1997).
However, as we have already emphasized, the global properties derived 
from equation~(6) are not affected by these ambiguities.

The sample of known DLA galaxies now includes about 100 objects.
Most of these have redshifts in the range $1.6 \la z \la 4$; only a
few are known at $z \la 1.6$ (which require observations from space)
and at $z \ga 4$ (which require bright quasars at higher redshifts).
We now have measurements of the column densities of neutral atomic 
hydrogen (HI) in all these systems and of the metals and dust in many 
of them 
(for HI, see Lanzetta, Wolfe, \& Turnshek 1995; Wolfe et al. 1995; 
Storrie-Lombardi, McMahon, \& Irwin 1996; Rao \& Turnshek 2000; 
Storrie-Lombardi \& Wolfe 2000; 
for metals, see Pettini et al. 1994, 1997a, 1999; Lu et al. 1996; 
Boiss\'e et al. 1998; Prochaska \& Wolfe 2000; 
for dust, see Pei, Fall, \& Bechtold 1991; Kulkarni, Fall, \& Truran 
1997; Pettini et al. 1997b; Welty et al. 1997; Vladilo 1998).
It is believed that the DLA galaxies have low fractions of ionized
hydrogen (HII) because the HI layers are opaque to ionizing radiation. 
In the few DLA galaxies that have been searched, the abundance of 
molecular hydrogen (H$_2$) lies below the value in the Milky Way 
(Levshakov et al. 1992; Ge \& Bechtold 1997).
One factor that potentially biases all these observations is the
presence of dust in DLA galaxies.
Quasars behind dusty DLA galaxies will be obscured, thus reducing
the chances that they will be included in optically selected samples 
(Ostriker \& Heisler 1984; Fall \& Pei 1993; Boiss\'e et al. 1998).
Gravitational lensing has the opposite effect, but this appears to 
be negligible in the existing samples (Le~Brun et al. 1997; Perna, 
Loeb, \& Bartelmann 1997; Smette, Claeskens, \& Surdej 1997).
Finally, we emphasize that estimates of the comoving densities of 
HI, metals, and dust are dominated by relatively few systems---those 
with the highest column densities.
As a result, they are less certain than is often recognized.

\section{PF Models}

Global evolution models of the type described above have been 
constructed by Lanzetta et al. (1995), Pei \& Fall (1995, hereafter 
PF), Malaney \& Chaboyer (1996), and Pei, Fall, \& Hauser (1999, 
hereafter PFH).
Some related issues are discussed by Songaila, Cowie, \& Lilly 
(1990), Timmes, Lauroesch, \& Truran (1995), Fall, Charlot, \& Pei 
(1996), Calzetti \& Heckman (1999), and Harwit (1999).
In the following, we present highlights from the PF and PFH models.
At the time the PF models were published, most of the successful
surveys for emission from galaxies were confined to $z \la 0.3$, and 
nothing was known about the global history of star formation, 
$\dot\Omega_s(z)$.
Indeed, one of the goals of the PF models was to predict 
$\dot\Omega_s(z)$ from absorption-line observations of DLA galaxies, 
particularly $\Omega_{\rm HI}(z)$ and $Z(z)$.
In 1994--95, the comoving density of HI was found to be a smoothly
increasing function of redshift from $z=0$ to $z\approx3$, while the 
mean metallicity was known at only two redshifts: $Z\approx Z_{\odot}$ 
at $z=0$ and $Z\approx0.1Z_{\odot}$ at $z\approx2$.
These observations---the first pertaining to gas consumption, the second 
to metal production---strongly suggested that most stars in the Universe
formed after $z\approx2$.
The motivation behind the global evolution models was to provide a 
framework to interpret such observations more quantitatively.

The PF models are based on the following simplifying assumptions:
1. HI is a tracer of the total interstellar content of galaxies,
both atomic and molecular ($\Omega_g\propto\Omega_{\rm HI}$). 
2. The dust content of galaxies is proportional to the metal content,
both locally and globally ($N_d\propto ZN_g$ and $\Omega_d\propto 
Z\Omega_g$).
3. The interaction between galaxies and the IGM takes one of three 
idealized forms: closed box ($\dot\Omega_f=0$), metal-free inflow 
($\dot\Omega_f=+\nu\dot\Omega_s$, $Z_f=0$), or metal-enriched outflow 
($\dot\Omega_f=-\nu\dot\Omega_s$, $Z_f=Z$).
Assumptions 1 and 2 are consistent with, and indeed motivated by, the
observations summarized in the previous section.
Assumption 2 couples the internal absorption and the obscuration of 
background quasars to the chemical evolution of galaxies; this enables 
self-consistent corrections for biases in the the census of DLA galaxies 
and hence in the estimates of $\Omega_{\rm HI}(z)$.
Assumption 3 is the cosmological analog of the standard treatment of
inflow and outflow in models of galactic chemical evolution, beginning
with the work of Larson (1972) and Hartwick (1976).
While one might question the precise validity of these assumptions,
they should be realistic enough to reveal a first glimpse of the global 
evolution of galaxies.

The PF models were designed to reproduce (as input) the observed
evolution in $\Omega_{\rm HI}(z)$.
As a bonus, they also matched (as output) the available estimates
of $Z(z)$ without any fine tuning of parameters.
Figure~2 shows the predicted evolution of the comoving rate of metal 
production $\dot\rho_Z(z)$ in the standard PF models.
(The results displayed in this and subsequent figures are for 
$H_0=50$~kms$^{-1}$Mpc$^{-1}$, $q_0=0.5$, and $\Lambda=0$.) 
This is just another way of expressing the star formation rate:
$\dot\rho_Z=y(3H_0^2/8\pi G)\dot\Omega_s$.
Figure~2 also shows subsequent emission-based estimates of 
$\dot\rho_Z(z)$ from several surveys, including the Canada-France 
Redshift Survey and the Hubble Deep Field (Gallego et al. 1995; 
Lilly et al. 1996; Madau et al. 1996, 1998; Connolly et al. 1997).
These were derived from rest-frame H$\alpha$ and UV emissivities and the 
approximate proportionality between UV emission and metal production in 
stellar populations.
Evidently, the predicted and observed rates are in broad agreement
(within factors of about two).
This is remarkable because the PF models were constructed only with 
absorption-line observations in mind, before the emission-based
estimates of $\dot\rho_Z$ became available.
The PF models, when combined with stellar population synthesis models,
also predicted a far-IR/sub-mm background (Fall et al. 1996) that was 
consistent with early empirical limits and estimates from the DIRBE 
and FIRAS experiments on {\it COBE} (Hauser 1996; Puget et al. 1996).

\begin{figure}
\plotfiddle{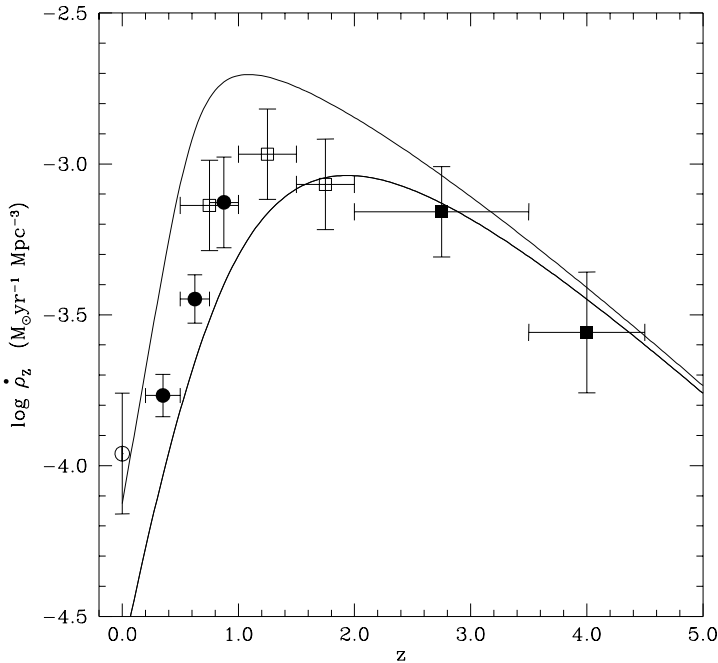}{3.5in}{0}{50}{50}{-160}{-95}
\caption{Comoving rate of metal production as a function of redshift.
The curves are from the PF global evolution models with the standard
parameters (upper curve with inflow, lower curve with closed-box or
outflow).
The data points correspond to cosmic H$\alpha$ and UV emissivities from
Gallego et al. (1995), Lilly et al. (1996), Madau et al. (1996, 1998),
and Connolly et al. (1997).}
\end{figure}

\section{PFH Models}

The PFH models were designed to improve upon the PF models by 
addressing several subsequent developments in this rapidly advancing 
field.
First, it was realized that estimates of the comoving star formation
and metal production rates from rest-frame UV emissivities were
potentially biased by absorption by dust.
The proposed corrections to the estimates of $\dot\rho_Z$ in Figure~2 
ranged from nearly nothing to more than an order of magnitude.
The large uncertainty in these corrections stemmed from the difficulty
of determining the dust content and wavelength dependence of the 
effective absorption in high-redshift galaxies.
Second, it was found from a large-area, ground-based survey that even 
the uncorrected UV emissivity at $z \ga 3$ was probably about twice
as high as had initially been estimated from the small-area Hubble 
Deep Field (Steidel et al. 1999).
Third, the limits on, and measurements of, the EBL intensity have 
improved, especially in the far-IR/sub-mm spectral region (Fixsen 
et al. 1998; Hauser et al. 1998; Schelgel, Finkbeiner, \& Davis 1998).
These are shown in Figure~3.
It now appears that there is at least as much energy in the 
long-wavelength hump of the EBL spectrum ($\lambda \ga 10$~$\mu$m) 
as in the short-wavelength hump ($\lambda \la 10$~$\mu$m).
This, of course, provides a valuable constraint on the amount of 
UV/optical emission that has been absorbed and reradiated by dust.

\begin{figure}
\plotfiddle{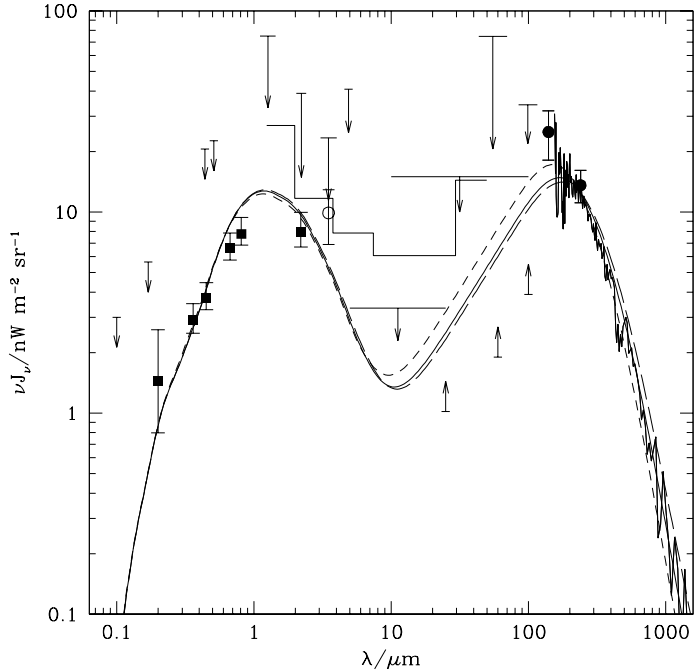}{3.5in}{0}{50}{50}{-160}{-95}
\caption{Extragalactic background intensity $J_\nu$ times frequency 
$\nu$ as a function of wavelength $\lambda$.  
The curves are from the PFH global evolution models.
References to the observations are given in the PFH paper.}
\end{figure}

In the PFH models, the observed emissivity at rest-frame UV 
wavelengths $E_{\nu}(z)$ is treated as an input function, along 
with $\Omega_{\rm HI}(z)$.
The emissivity at other wavelengths is computed from stellar 
population synthesis models and a template spectrum of dust emission 
derived from the local emissivity at several IR wavelengths. 
The absorption per mass of dust, assumed proportional to the mass
of interstellar metals, is then adjusted to satisfy the empirical 
constraints on the EBL intensity.
This procedure closes the equations of cosmic chemical evolution 
without invoking a specific relation between the star formation 
rate and the inflow or outflow rate, as in the PF models.
Indeed, in the PFH models, both $\dot\Omega_s(z)$ and 
$\dot\Omega_f(z)$ are independent output solutions of the equations.
The net effect of this procedure is to determine the correction for
absorption by dust between the observed and true UV emissivities, i.e., 
between the observed and true star formation rates, as a function of 
redshift, in a way that is consistent with the production of metals 
and dust and the consumption or replenishment of gas by a combination 
of star formation and inflow or outflow.
The results are also guaranteed, by construction, to be consistent
with a wide variety of observations, including the UV emissivity 
and comoving density of HI as functions of redshift and the relative
amounts of energy in the long- and short-wavelength humps of the
EBL spectrum.

\begin{figure}
\plotfiddle{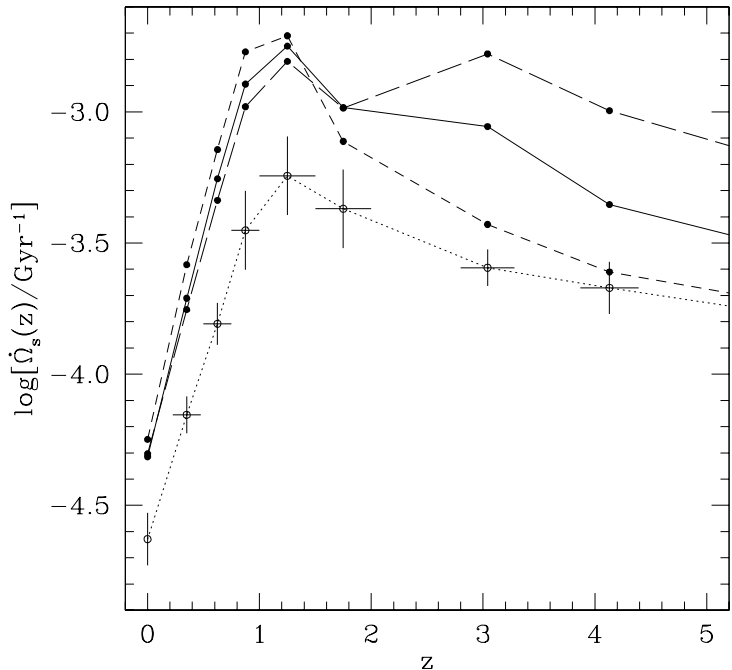}{3.5in}{0}{50}{50}{-160}{-95}
\caption{Comoving rate of star formation as a function of redshift.
The open circles represent the observed rate, based on rest-frame 
UV emissivities, from Steidel et al. (1999), while the filled circles
represent the rate corrected for absorption by dust in the PFH models.}
\end{figure}

As Figure~3 shows, the EBL intensity in the models does indeed match
the observations. 
(In Figures 3--5, the short-dash, solid, and long-dash lines
indicate the range of solutions permitted by uncertainties in 
$J_\nu$ at sub-mm wavelengths.)
Figure~4 shows the evolution of the comoving rate of star formation 
$\dot\Omega_s(z)$ before and after the corrections for absorption by 
dust.
Evidently, these corrections are factors of 2--4 at most redshifts.
The star formation rates in the PFH models have the same qualitative
behavior as those in the PF models, including the rapid rise from 
$z=0$ to $z=$~1--2, but the decline from $z=$~1--2 to $z\approx4$ 
is now shallower or possibly non-existent.
Figure~5 shows the predicted evolution of the mean interstellar
metallicity and the comoving density of interstellar metals,
$Z(z)$ and $\Omega_m^{ism}(z)$.
The former increases monotonically with decreasing redshift, while
the latter first increases and then decreases as the interstellar
medium is consumed by star formation.
The effective yield in the models has been adjusted to give $Z = 
Z_{\odot}$ at $z = 0$, the approximate mean interstellar metallicity 
in present-day galaxies (averaged over internal gradients and the 
luminosity function of galaxies).
At higher redshifts, the mean metallicities in the models are 
marginally consistent with those in the DLA galaxies, although the 
observed enrichment appears to be slower than predicted.
This may in part be a consequence of the bias against selecting dusty 
and hence metal-rich DLA galaxies, an effect that tends to make the
observed $Z(z)$ relation artificially flat.

\begin{figure}
\plotfiddle{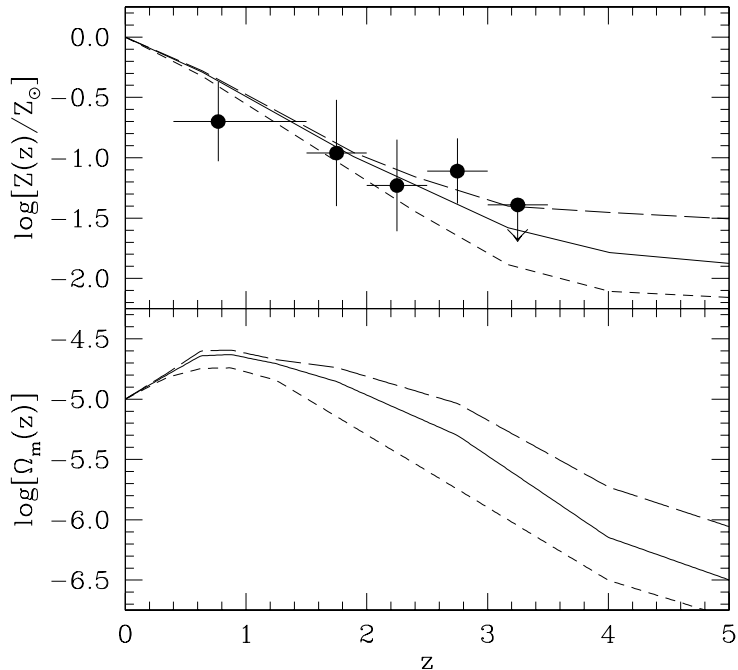}{3.5in}{0}{50}{50}{-160}{-95}
\caption{Mean interstellar metallicity in units of the solar value 
(upper panel) and comoving density of interstellar metals (lower 
panel) as functions of redshift.
The curves are from the PFH global evolution models.
The data point at $z=0.8$ is from Boiss\'e et al. (1998), while those
at $z\ge1.7$ are from Pettini et al. (1997a).}
\end{figure}

\section{Conclusions}

The main conclusion to be drawn from this article is that we now 
have the apparatus needed to interpret simultaneously a wide variety 
of observations pertaining to the average properties of galaxies, 
including the comoving densities of stars and interstellar gas, 
metals, and dust.
An attraction of the global evolution models is that they relate 
observations that would otherwise appear disparate, namely, those of 
the light galaxies emit and those of the light they absorb (from 
background quasars).
Moreover, this approach is perhaps the simplest and most natural one 
in which to interpret the EBL.
Simplicity is achieved by focusing on the average properties of
galaxies and, for the most part, ignoring their individuality and 
internal complexity.

The global evolution models have now been developed to the point that 
the accuracy of the results is limited more by observational than by 
theoretical uncertainties.
The PFH models incorporated the best data available in 1998.
Since then, there have been debates about the evolution of the 
comoving density of HI and the mean metallicity in DLA galaxies.
The evolution of the UV emissivity, and hence the comoving rate 
of star formation, is probably equally uncertain, even before the
corrections for absorption by dust.
The problem is that all these observational relations are based on 
relatively small samples and are affected by a variety of selection 
biases (e.g., dust for absorption, surface brightness for emission).
Thus, it should be a high priority in the next few years to improve 
the observations, by increasing the sizes of the samples (e.g., by 
factors of 10 or more) and by reducing the selection biases (e.g., 
by using radio rather than optically selected quasars).
It should then be possible to repeat much of the analysis described
here with greater confidence in the accuracy of the results.

\end{document}